\documentclass[letterpaper]{article} 
\usepackage{aaai25}  
\usepackage{times}  
\usepackage{helvet}  
\usepackage{courier}  
\usepackage[hyphens]{url}  
\usepackage{graphicx} 
\usepackage{natbib}  
\usepackage{caption} 
\usepackage{amsmath} 
\usepackage{amssymb}
\usepackage{multirow}
\usepackage{colortbl}
\usepackage{graphicx}
\usepackage{fancyhdr}
\usepackage[dvipsnames]{xcolor}
\usepackage{pdfpages}
\usepackage{booktabs} 
\usepackage{appendix}
\usepackage{algorithm}
\usepackage{algorithmic}

\usepackage{newfloat}
\usepackage{listings}
\DeclareCaptionStyle{ruled}{labelfont=normalfont,labelsep=colon,strut=off} 
\floatstyle{ruled}
\newfloat{listing}{tb}{lst}{}
\floatname{listing}{Listing}
\title{BeFA: A General Behavior-driven Feature Adapter for \\ Multimedia Recommendation}
\author {
    Qile Fan\textsuperscript{\rm 1} \equalcontrib,
    Penghang Yu\textsuperscript{\rm 1} \equalcontrib,
    Zhiyi Tan\textsuperscript{\rm 1},
    Bing-Kun Bao\textsuperscript{\rm 3},
    Guanming Lu\textsuperscript{\rm 1, 2} \thanks{Corresponding author.}
}
\affiliations {
    \textsuperscript{\rm 1}{School of Communications and Information Engineering, \\Nanjing University of Posts and Telecommunications, Nanjing, China}\\
    \textsuperscript{\rm 2}{Jiangsu Key Laboratory of Intelligent Information Processing and Communication Technology, Nanjing, China}\\
    \textsuperscript{\rm 3}{School of Computer Science, Nanjing University of Posts and Telecommunications, Nanjing, China}\\
    b21011331@njupt.edu.cn, 2022010201@njupt.edu.cn, tzy@njupt.edu.cn, bingkunbao@njupt.edu.cn, lugm@njupt.edu.cn 
}

\usepackage{bibentry}

\begin{document}

\maketitle

\begin{abstract}
Multimedia recommender systems focus on utilizing behavioral information and content information to model user preferences. Typically, it employs pre-trained feature encoders to extract content features, then fuses them with behavioral features. However, pre-trained feature encoders often extract features from the entire content simultaneously, including excessive preference-irrelevant details. We speculate that it may result in the extracted features not containing sufficient features to accurately reflect user preferences.
To verify our hypothesis, we introduce an attribution analysis method for visually and intuitively analyzing the content features. The results indicate that certain items' content features exhibit the issues of \textbf{information drift} and \textbf{information omission}, reducing the expressive ability of features. Building upon this finding, we propose an effective and efficient general \textbf{Be}havior-driven \textbf{F}eature \textbf{A}dapter (\textbf{BeFA}) to tackle these issues. This adapter reconstructs the content feature with the guidance of behavioral information, enabling content features accurately reflecting user preferences. Extensive experiments demonstrate the effectiveness of the adapter across all multimedia recommendation methods. Our code is made publicly available on \url{https://github.com/fqldom/BeFA}.
\end{abstract}
%

\section{Introduction}

Recommender systems have gained widespread adoption across various domains, aiming to assist users in discovering information that aligns with their preferences \cite{vldb1,vldb2}. In the case of multimedia platforms, the abundance of data resources provides recommender systems with increased opportunities to accurately model user preferences \cite{survey}.

Existing multimedia recommendation methods typically involves two main steps \cite{w2go,survey}. First, a pre-trained feature encoder is employed to capture content features from diverse modalities. Subsequently, these content features are fused with behavioral features to obtain user preference representations. Recently, researchers have focused on enhancing the quality of these representations through self-supervised learning. For instance, SLMRec \cite{slmrec} investigates potential relationships between modalities through the use of contrastive learning, thereby obtaining a powerful representation. BM3 \cite{bm3} utilizes a dropout strategy to construct multiple views and reconstructs the interaction graph, incorporating intra- and inter-modality contrastive loss to facilitate effective representation learning. MICRO \cite{micro} and MGCN \cite{mgcn} maximize the mutual information between content features and behavioral features with a self-supervised auxiliary task, and have achieved excellent performance. 
\begin{figure}[t!]
	\centering
	\includegraphics[width=\linewidth]{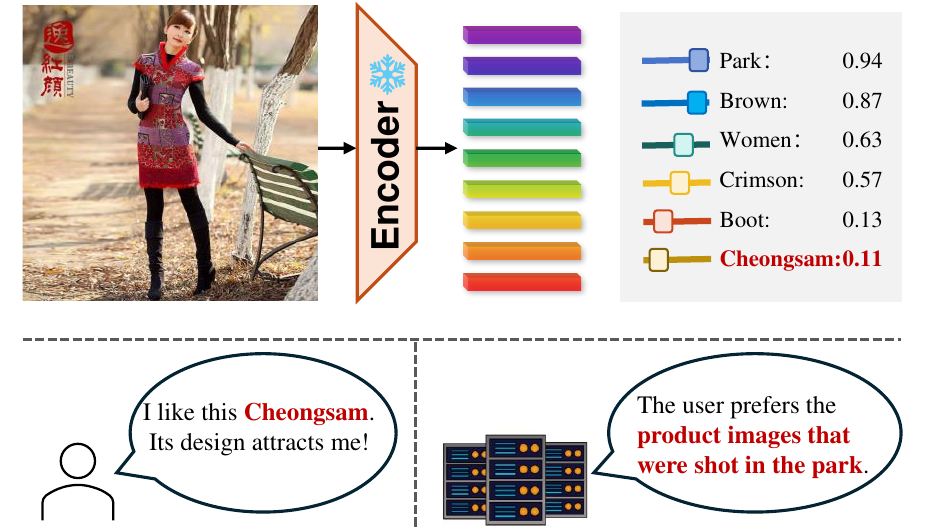} 
	\caption{Illustration of content features that do not accurately reflect the users' preferences. Excessive irrelevant information hinders the recommender system's ability to effectively model the users' true preferences.}
	\label{fig:introduction}
    \vspace{-10pt}
\end{figure}
\begin{figure*}[t!]
	\centering
	\includegraphics[width=\textwidth]{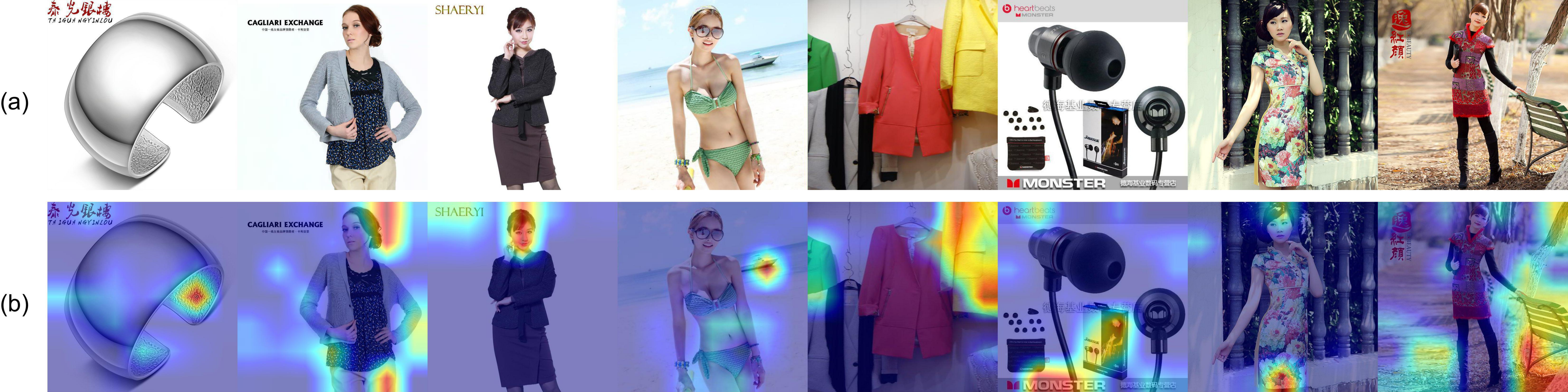} 
	\caption{Results of visualisation attribution analysis on the TMALL dataset. The  first row contains the original item images while the second row displays the corresponding heatmaps. The four samples on the left reflect information drift and the four samples on the right reflect information omission.}
	\label{deficiency_heatmap}
\end{figure*}
Despite the good progress made in utilizing content and behavioral information more effectively, a crucial yet easily overlooked problem arises: \textbf{Are content features obtained from pre-trained encoders containing sufficient features to reflect user preferences?} Intuitively, multimedia content inherently exhibits the characteristic of low informational value density, where a significant portion of the presented information may be irrelevant to the users' focus \cite{survey}. Pre-trained feature encoders extract information from the entire content simultaneously, which can result in content features that do not truly reflect the users' preferences (as shown in Figure \ref{fig:introduction}). Fusing these irrelevant content features with behavioral features may mislead user preference modeling, resulting in suboptimal recommendation performance.

To answer this question, we introduce a similarity-based attribution analysis method for visualizing and intuitively analyzing the content features. This method evaluates the extent to which content features can reflect user preferences, enabling researchers to visually assess the quality of content features for the first time. The results indicate that not all items' content features accurately reflect user preferences. Due to the presence of irrelevant information, certain items' content features exhibit the issues of \textbf{information drift} and \textbf{information omission}. As shown in Figure \ref{deficiency_heatmap}, some items' content features do not include information about the items that users are interested in, but instead erroneously include information about unrelated items, a phenomenon we term \textbf{information drift}. There are also some items' content features that omit certain key details of the items, which we refer to as \textbf{information omission}. These issues ultimately prevent recommender systems from accurately modeling user preferences. Furthermore, we propose a plug-and-play general \textbf{Be}havior-driven \textbf{F}eature \textbf{A}dapter (\textbf{BeFA}) to address the discovered issues. This adapter effectively decouples, filters and reconstructs content features, leveraging behavioral information as a guide to obtain more precise representations of content information. Extensive experiments demonstrate the adapter's effectiveness across various recommendation methods and feature encoders.

Our main contributions can be summarized as follows:
\begin{itemize}
  \item We introduce a similarity-based visual attribution method, which enables researchers to visually analyze the quality of content features for the first time.
  \item We experimentally revealed the issues of information drift and information omission in content features.
  \item We propose a general behavior-driven feature adapter, which obtain more precise content representation through decoupling and reconstructing content features.
\end{itemize}
\section{Related Work}
\begin{figure*}[t]
	\centering
	\includegraphics[width=\textwidth]{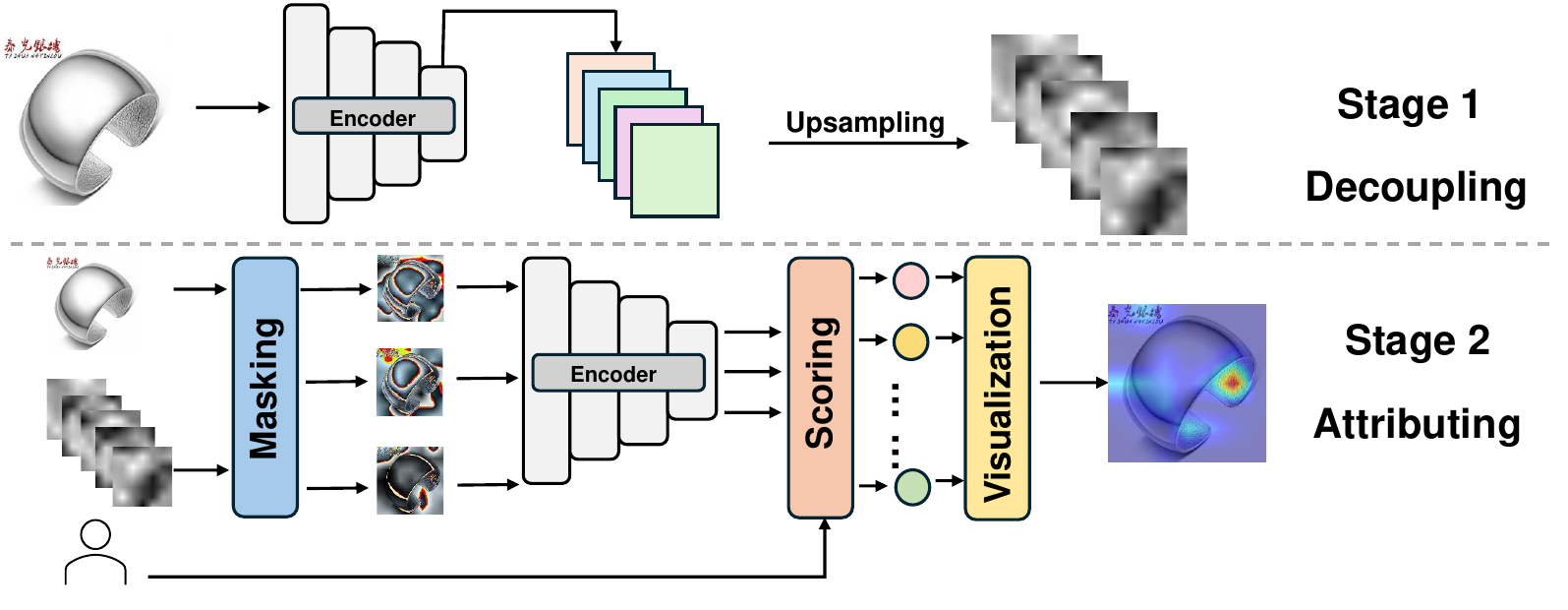} 
	\caption{Pipeline of the proposed attribution analysis method.}
	\label{analysis pipeline}
\end{figure*}
\subsection{Multimedia Recommendation}
Collaborative filtering (CF) based methods leverage behavioral similarities for top-k recommendations \cite{cf1}. To improve the performance of CF-based methods, researchers integrate item multimodal content. Typically, they employ pre-trained neural networks to extract content features, and then fuse content features with behavioral features to more effectively model user preference. \cite{w2go}. For instance, MMGCN \cite{mmgcn} creates modality-specific interaction graphs. LATTICE \cite{lattice} adds links between similar items. FREEDOM \cite{FREEDOM} constructs graphs using sensitivity edge pruning. Recently, self-supervised learning methods like BM3 \cite{bm3} and MGCN \cite{mgcn} further enhance representation quality by reconstructing interaction graphs and adaptively learning content feature importance. However, item content information often contains much irrelevant information and exhibiting low value density. We argue pre-extracted item content features insufficiently reflect users' preferences. Directly incorporating these features may mislead user preference modeling, resulting in suboptimal recommendation performance.
\subsection{Attribution Analysis}
Modality feature attribution analysis methods such as Class Activation Mapping (CAM) \cite{CAM}, Grad-CAM \cite{gradcam}, and Score-CAM \cite{scoreCAM} generate heatmaps to visualize the decision-making process of convolutional neural networks. Grad-CAM uses gradients for better applicability across network architectures. Score-CAM weight activation maps with output scores, which eliminates the dependence on gradients.
However, current attribution analysis methods primarily focus on analyzing the encoder itself and fail to visually analyze items' content features in recommender systems. Most of these methods use gradient information to obtain weights, which reflect more of the encoder focus. As a result, they fail to capture the similarity between content features and behavioral features, which directly reflect the user's preferences. Consequently, these methods are unsuitable for use in recommender systems. Thus, to better analyze the quality of item content features, we need to design a novel similarity-based  attribution analysis method.
 
\subsection{Parameter-Efficient Adaptation}
In multimodal domains, semantic differences limit feature expressiveness, impacting model performance \cite{nlp1,cv1}. Using pre-trained encoders to extract features in real-world scenarios can result in not accurately reflecting user preferences. To address this challenge, researchers have explored techniques such as fine-tuning and adapters to reduce the impact of semantic differences. In the context of recommendation scenarios where data is continuously changing and expanding, performing full fine-tuning of the encoder faces huge challenges. Due to the above limitations, we consider low-parameter fine-tuning methods to efficiently adapt content features in real time. Low-Rank Adaptation (LoRA) \cite{lora} integrate low-rank decomposition matrices, while Prompt Tuning \cite{softprompt} uses learnable embedding vectors as hints. 
However, adding adapters to the middle layer of the pre-trained model may lead to information loss. Existing methods are not generalized for recommendation tasks. They fail to adequately incorporate behavioral information, which is crucial for intuitively reflecting users' preferences.
\section{Preliminaries}

\subsection{Attribution Analysis}
Existing attribution methods are mainly focused on pre-trained encoders, which can only analyze the quality of the content features themselves, fail to reflect their contributions to recommendations. Considering that behavioral features directly reflect user preferences in recommendation, we propose a similarity-based attribution analysis. This method explores the quality of content features extracted by a pre-trained encoder in recommendations. By calculating the cosine similarity of each pixel to the behavioral features, we generate a heatmap that visually identifies the content features relevant to users' behaviors.

We follow the core principle of Class Activation Mapping (CAM) to linearly weight feature maps. The main difference compared to the previous series of CAM methods is the way in which the linear weights are obtained. Our approach consists of two primary stages. In the first stage, we take the item image $I$ as input to the corresponding encoder, obtaining $N$ images on the corresponding channels at the target layer denoted as ${a_l}$.  Then we upsample it, denoted as $a_l^{up}$. This is done to generate masked images corresponding to different parts. Considering in all types of encoders, layers closer to the prediction layer contain the highest level and most abstract information \cite{attention}. Consequently, for the ResNet family visual encoder, we select the last ReLU layer of the final Bottleneck as the target layer, while for the ViT family visual encoder, we choose the penultimate ResidualAttentionBlock. The corresponding content feature $\boldsymbol{x_l}$ is extracted from each mask $a_l^{up}$. Behavioural features $\boldsymbol{h_i}$ are obtained from existing multimedia recommendation models. The weights of the corresponding content features are obtained by calculating the similarity between them.
\begin{figure*}[t]
	\centering
	\includegraphics[width=0.98\textwidth]{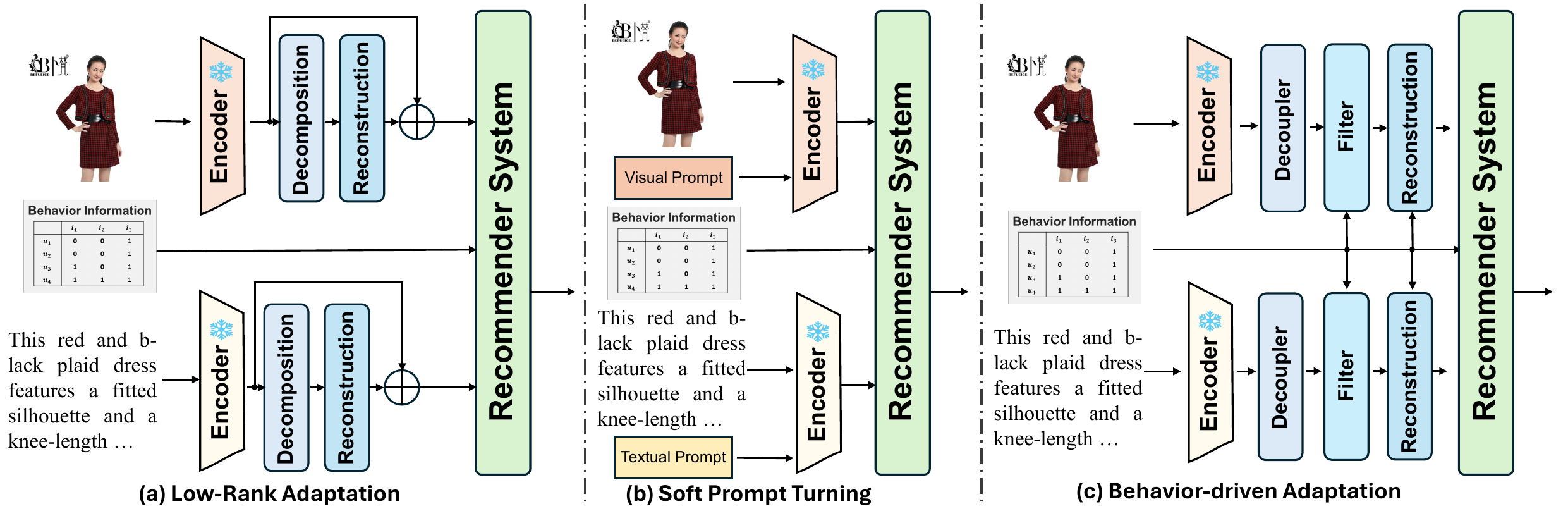} 
	\caption{Comparison of parameter tuning methods. \textbf{(a)} Low-Rank Adaptation  \textbf{(b)} Soft Prompt Turning  \textbf{(c)} BeFA.}
	\label{adaptations}
\end{figure*}
\begin{equation}
  \text{similarity}(\boldsymbol{x_l}, \boldsymbol{h_i}) = \frac{{\boldsymbol{x_l} \cdot \boldsymbol{h_i}}}{{\|\boldsymbol{x_l}\| \cdot \|\boldsymbol{h_i}\|}}
\end{equation}
Ultimately, the final visualization result is achieved by linearly weighting and summing the feature maps with the similarity. The heatmap $M$ is represented as:
\begin{equation}
    M=\sum_{l}^N a_l\times \text{similarity}(f_{pre}(I_c \odot a_l^{up}),\boldsymbol{h_i})
\end{equation}
where $\odot$ is the Hadamard product. $I_c$ are the all color channels of the image. $f_{pre}$ denotes the process of extracting features using the corresponding encoder.
\subsection{Deficiency Analysis}
From Figure \ref{deficiency_heatmap} we observed that the pre-trained feature encoder suffers from the issues of wrong region of interest (information drift) and insufficient region of interest (information omission). Specifically, although the user wants to buy clothes, the feature encoder incorrectly focuses on the face of the person who wears the cloth, a phenomenon we call \textbf{information drift}. Meanwhile, we also found that some of the features did not fully reflect all the characteristics of the item. Such as only focusing on certain details of the clothes and failing to focus on its whole, a phenomenon we call \textbf{information omission}. Information drift can cause the recommender system to ignore useful and critical information. This leads to a loss of important guidance for the recommendation task. Which lead a mismatch between recommendations and users' preferences. Information omission can result in important visual or textual cues being ignored or incomplete. This causes a loss of critical information needed for the recommendation task. These issues compromise the overall performance of the recommender system. As a result, directly applying the  pre-extracted content features will inevitably impair the performance of the recommender system. In the Appendix, we perform a theoretical analysis to further illustrate the point.

\section{Feature Adapter}
To address the discovered issues, we propose Behavior-driven Feature Adapter (BeFA). The purpose of this adapter is to efficiently adapt content features to improve their quality in recommendations, thus improving the performance of recommendations. Specifically, the content features are first extracted by a pre-trained encoder and then adapted by BeFA. Then it is fed into the multimodal recommender system for item and user modeling. BeFA and the downstream recommender system share the optimization objective and adopt End-to-End training strategy.
\subsection{Problem Formulation}
\begin{table*}[t]
  \centering
  \small
  \setlength{\tabcolsep}{0.75mm} 
  \setlength{\abovecaptionskip}{0.05cm}
  \begin{tabular*}{\linewidth}{c|c|cccc|cccc|cccc}
      \toprule
      & \multicolumn{1}{c|}{}                           & \multicolumn{4}{c|}{TMALL}                                                                                            & \multicolumn{4}{c|}{Microlens}                                                                                       & \multicolumn{4}{c}{H\&M}                                                                                               \\
\multirow{-2}{*}{Encoder}    & \multicolumn{1}{c|}{\multirow{-2}{*}{Datasets}} & {\color[HTML]{467886} R@10} & {\color[HTML]{467886} R@20} & {\color[HTML]{467886} N@10} & {\color[HTML]{467886} N@20} & {\color[HTML]{467886} R@10} & {\color[HTML]{467886} R@20} & {\color[HTML]{467886} N@10} & {\color[HTML]{467886} N@20} & {\color[HTML]{467886} R@10} & {\color[HTML]{467886} R@20} & {\color[HTML]{467886} N@10} & {\color[HTML]{467886} N@20} \\ \cline{1-14}
      & BM3                                             & 0.0189                      & 0.0298                      & 0.0102                      & 0.0132                      & 0.0510                      & 0.0851                      & 0.0278                      & 0.0375                      & 0.0204                      & 0.0320                      & 0.0114                       & 0.0144                      \\
      & BM3+BeFA                                        & 0.0212                      & 0.0319                      & 0.0115                      & 0.0144                      & 0.0566                      & 0.0911                      & 0.0314                      & 0.0412                      & 0.0266                      & 0.0391                      & 0.0157                       & 0.0190                      \\
      & \textit{Improve}                                & 12.17\%                     & 7.05\%                      & 12.75\%                     & 9.09\%                      & 10.98\%                     & 7.05\%                      & 12.95\%                     & 9.87\%                      & 30.39\%                     & 22.19\%                     & 37.72\%                      & 31.94\%                     \\ \cline{2-14}
      & LATTICE                                         & 0.0238                      & 0.0356                      & 0.0134                      & 0.0167                      & 0.0553                      & 0.0886                      & 0.0308                      & 0.0402                      & 0.0289                      & 0.0427                      & 0.0161                       & 0.0197                      \\
      & LATTICE+BeFA                                    & 0.0260                      & 0.0403                      & 0.0183                      & 0.0183                      & 0.0593                      & 0.0943                      & 0.0328                      & 0.0427                      & 0.0317                      & 0.0498                      & 0.0171                       & 0.0217                      \\
      & \textit{Improve}                                & 9.24\%                      & 13.20\%                     & 36.57\%                     & 9.58\%                      & 7.23\%                      & 6.43\%                      & 6.49\%                      & 6.22\%                      & 9.69\%                      & 16.63\%                     & 6.21\%                       & 10.15\%                     \\ \cline{2-14} 
      & FREEDOM                                         & 0.0212                      & 0.0340                      & 0.0113                      & 0.0148                      & 0.0474                      & 0.0774                      & 0.0262                      & 0.0348                      & 0.0348                      & 0.0526                      & 0.0188                       & 0.0234                      \\
      & FREEDOM+BeFA                                    & 0.0253                      & 0.0375                      & 0.0136                      & 0.0170                      & 0.0503                      & 0.0814                      & 0.0279                      & 0.0368                      & 0.0409                      & 0.0583                      & 0.0226                       & 0.0271                      \\
      & \textit{Improve}                                & 19.34\%                     & 10.29\%                     & 20.35\%                     & 14.86\%                     & 6.12\%                      & 5.17\%                      & 6.49\%                      & 5.75\%                      & 17.53\%                     & 10.84\%                     & 20.21\%                      & 15.81\%                     \\ \cline{2-14} 
            & MGCN                                      & 0.0249                      & 0.0380                      & 0.0135                      & 0.0171                      & 0.0618                      & 0.0972                      & 0.0342                      & 0.0442                      & 0.0367                      & 0.0549                      & 0.0204                       & 0.0251                      \\
      & MGCN+BeFA                                       & 0.0261                      & 0.0395                      & 0.0142                      & 0.0179                      & 0.0630                      & 0.1000                      & 0.0351                      & 0.0456                      & 0.0405                      & 0.0594                      & 0.0225                       & 0.0274                      \\
      & \textit{Improve}                                & 4.82\%                      & 3.95\%                      & 5.19\%                      & 4.68\%                      & 1.94\%                      & 2.88\%                      & 2.63\%                      & 3.17\%                      & 10.35\%                     & 8.20\%                      & 10.29\%                      & 9.16\%                      \\ \cline{2-14} 
    \rowcolor{gray!40} 
     \multirow{-13}{*}{CLIP}      
     & \textit{\textbf{Avg Improve}}              & \textbf{11.39\%}                       & \textbf{8.62\%}                        & \textbf{18.71\%}                     & \textbf{9.55\%}                      & \textbf{6.57\%}                        & \textbf{5.38\%}                        & \textbf{7.14\%}                      & \textbf{6.25\%}                      & \textbf{16.99\%}                       & \textbf{14.46\%}                       & \textbf{18.61\%}                     & \textbf{16.77\%}                     \\ \cline{1-14}
      & BM3                                             & 0.0184                      & 0.0299                      & 0.0097                      & 0.0129                      & 0.0508                      & 0.0842                      & 0.0279                      & 0.0373                      & 0.0195                      & 0.0304                      & 0.0107                       & 0.0135                      \\
      & BM3+BeFA                                        & 0.0224                      & 0.0322                      & 0.0125                      & 0.0152                      & 0.0537                      & 0.0877                      & 0.0299                      & 0.0395                      & 0.0248                      & 0.0378                      & 0.0149                       & 0.0183                      \\
      & \textit{Improve}                                & 21.74\%                     & 7.69\%                      & 28.87\%                     & 17.83\%                     & 5.71\%                      & 4.16\%                      & 7.17\%                      & 5.90\%                      & 27.18\%                     & 24.34\%                     & 39.25\%                      & 35.56\%                     \\   \cline{2-14}
      & LATTICE                                         & 0.0252                      & 0.0374                      & 0.0139                      & 0.0173                      & 0.0580                      & 0.0953                      & 0.0320                      & 0.0426                      & 0.0293                      & 0.0439                      & 0.0164                       & 0.0202                      \\
      & LATTICE+BeFA                                    & 0.0266                      & 0.0403                      & 0.0147                      & 0.0184                      & 0.0633                      & 0.1021                      & 0.0340                      & 0.0451                      & 0.0316                      & 0.0498                      & 0.0171                       & 0.0218                      \\
      & \textit{Improve}                                & 5.56\%                      & 7.75\%                      & 5.76\%                      & 6.36\%                      & 9.14\%                      & 7.14\%                      & 6.25\%                      & 5.87\%                      & 7.85\%                      & 13.44\%                     & 4.27\%                       & 7.92\%                      \\ \cline{2-14} 
      & FREEDOM                                         & 0.0197                      & 0.0319                      & 0.0107                      & 0.0140                      & 0.0613                      & 0.0976                      & 0.0337                      & 0.0440                      & 0.0364                      & 0.0553                      & 0.0192                       & 0.0241                      \\
      & FREEDOM+BeFA                                    & 0.0259                      & 0.0377                      & 0.0141                      & 0.0174                      & 0.0641                      & 0.1004                      & 0.0356                      & 0.0459                      & 0.0417                      & 0.0613                      & 0.0234                       & 0.0285                      \\
      & \textit{Improve}                                & 31.47\%                     & 18.18\%                     & 31.78\%                     & 24.29\%                     & 4.57\%                      & 2.87\%                      & 5.64\%                      & 4.32\%                      & 14.56\%                     & 10.85\%                     & 21.88\%                      & 18.26\%                     \\ \cline{2-14} 
      & MGCN                                            & 0.0266                      & 0.0403                      & 0.0144                      & 0.0182                      & 0.0693                      & 0.1075                      & 0.0388                      & 0.0496                      & 0.0405                      & 0.0613                      & 0.0218                       & 0.0272                      \\
      & MGCN+BeFA                                       & 0.0275                      & 0.0414                      & 0.0152                      & 0.0185                      & 0.0702                      & 0.1085                      & 0.0389                      & 0.0498                      & 0.0451                      & 0.0655                      & 0.0246                       & 0.0299                      \\
      & \textit{Improve}                                & 3.38\%                      & 2.73\%                      & 5.56\%                      & 1.65\%                      & 1.30\%                      & 0.93\%                      & 0.26\%                      & 0.40\%                      & 11.36\%                     & 6.85\%                      & 12.84\%                      & 9.93\%                      \\ \cline{2-14}
      \rowcolor{gray!40} 
      \multirow{-13}{*}{ImageBind} 
      & \textit{\textbf{Avg Improve}}                   & \textbf{15.54\%}                       & \textbf{9.09\%}                        & \textbf{17.99\%}                     & \textbf{12.53\%}                     & \textbf{5.18\%}                        & \textbf{3.77\%}                        & \textbf{4.83\%}                      & \textbf{4.12\%}                      & \textbf{15.24\%}                       & \textbf{13.87\%}                       & \textbf{19.56\%}                     & \textbf{17.92\%}  \\

  \bottomrule
\end{tabular*}
  \caption{Performance Comparison on Different Recommender Models. The $t$-tests validate the significance of performance improvements with $p$-value $\leq$ 0.05.}
   \label{performance}
\end{table*}

Let $u\in \mathcal{U}$ and $i\in \mathcal{I}$ denote the user and item, respectively. The input behavioral features for user $u$ and item $i$ are represented as $ \mathrm{E_{b}} \in \mathbb{R}^{d \times (|\mathcal{U}|+|\mathcal{I}|)} $, where $ d $ is the embedding dimension. Each row represents the user embedding $\boldsymbol{h_u}$ and item embedding $\boldsymbol{h_i}$. They are initialised with ID information \cite{vbpr}. Each item content feature is denoted as $ \boldsymbol{e_{i,m}} \in \mathbb{R}^{d_m} $, where $ d_m $ is the dimension of the features. $ m \in \mathcal{M} $ represents the modality and $ \mathcal{M} $ is the set of modalities. Multimedia recommendation aims to rank items for each user by predicting preference scores $ \hat{y}_{u,i}$, which indicates the possibility of interaction between the user $u$ and the item $i$. The recommended process can be viewed as a function $f(\cdot)$. This process can be viewed simply as:
\begin{equation}
    \hat{y}_{u,i} = {f}(\boldsymbol{h_u},\boldsymbol{h_i},\boldsymbol{e_{i,m}})
\end{equation}
Existing multimodal recommendation methods typically use pre-trained encoders to extract content features $\boldsymbol{e_{i,m}}$. Due to the deficiencies of the pre-trained encoder, a large deviation in the content features $\boldsymbol{e_{i,m}}$ leads to a bias in the prediction score $\hat{y}_{u,i}$, which affects the recommendation results. Therefore, we designed BeFA to adapt the content features to make them more suitable for recommendation. BeFA improves the recommendation results by adjusting $\boldsymbol{e_{i,m}}$, which can be viewed as a function $g(\cdot)$. The adjusted predicted preference scores is donated:
\begin{equation}
    \hat{y}_{u,i} = {f}[\boldsymbol{h_u},\boldsymbol{h_i},g(\boldsymbol{h_i},\boldsymbol{e_{i,m}})]
\end{equation}
\subsection{Behavior-driven Feature Adapter}
Due to the shortcomings of pre-trained feature encoders, their extracted content features contain amount of irrelevant and erroneous information. To better utilize modality information,we propose \textbf{Be}haviour-driven \textbf{F}eature \textbf{A}dapter(\textbf{BeFA}) for adapting content features. Firstly, we decouple the original item $i$ content features $\boldsymbol{e_{i,m}}$ in the decoupled feature space:
\begin{equation}
  \dot{\boldsymbol{e}}_{i,m}=\mathbf{W}_1\boldsymbol{e}_{i,m}+\mathbf{b}_1,
\end{equation}
where $\mathbf{}{W}_1\in\mathbf{}{R}^{d_{m}\times d_{a}}$ and $\mathbf{b}_1\in\mathbb{R}^{d_a}$ represent trainable transformation matrix and bias vector. $d_{a}$ is a hyper-parameter represents the dimension of the decoupled space.

Considering that the behavioral information reflects the users' preference \cite{w2go}, we filter the preference-related content features with the guidance of behavioral information. The filter function $f_{gate}^{m}(\cdot)$ is represented as:
\begin{equation}
  \ddot{\boldsymbol{e}}_{i,m}=f_{gate}^{m}(\boldsymbol{b}_{i},\dot{\boldsymbol{e}}_{i,m})=\boldsymbol{b}_{i}\odot\tanh(\mathbf{W}_{2}\dot{\boldsymbol{e}}_{i,m}+\mathbf{b}_{2}),
\end{equation}
where $\mathbf{W}_2\in\mathbb{R}^{d_{a}\times d_{a}}$ and $\mathbf{b}_2\in\mathbb{R}^{d_a}$ are trainable parameters, $\boldsymbol{b_i}$ denotes the behavioral information of item i. Here $\tanh(\cdot)$ is the Tanh non-linear transformation.

Finally, the decoupled content features are selectively recombined. By combining different decoupled features with behavioral guidance, we aim to enhance the system's ability to capture nuanced and contextually relevant aspects of the content. The combination function $f_{merge}^{m}$ is represented as:
\begin{equation}
  \overline{\boldsymbol{e}}_{i,m}=f_{merge}^{m}(\boldsymbol{b}_{i},\ddot{\boldsymbol{e}}_{i,m})=\boldsymbol{b}_{i}\odot\sigma(\mathbf{W}_{3}\ddot{\boldsymbol{e}}_{i,m}+\mathbf{b}_{3}),
\end{equation}
where $\mathbf{W}_3\in\mathbb{R}^{d_{a}\times d_{m}}$ and $\mathbf{b}_3\in\mathbb{R}^{d_m}$ are trainable parameters. $\sigma(\cdot)$ is the sigmoid non-linear transformation. It is important to note that although Equations 6 and 7 are formally similar, they achieve different effects.

Moreover, we introduce ReLU activation functions and Dropout between each transformation layer. This enables the model to learn non-linear relationships, thereby improving its fitting ability and expressive power. Additionally, it enhances the model's generalization, making it more suitable for practical application scenarios.

\section{Experiments}

\subsection{Experimental Settings}
\subsubsection{Datasets}

We conducted experiments on three publicly available datasets: (a) {TMALL}\footnote{\url{https://tianchi.aliyun.com/dataset/140281}}; (b) {Microlens}\footnote{\url{https://recsys.westlake.edu.cn/MicroLens-50k-Dataset/}};  and (c) {H\&M}\footnote{\url{https://www.kaggle.com/datasets/odins0n/handm-dataset-128x128}}. The detailed information is presented in the Appendix. For multimodal information, we utilized pre-trained CLIP \cite{clip} and ImageBind \cite{imagebind} to extract aligned multimodal features. 
\subsubsection{Comparative Model Evaluation}
To verify the prevalence of feature encoder defects, we used ViT-B/32-based CLIP and ImageBind for our experiments. CLIP and ImageBind are advanced cross-modality feature encoders. CLIP known for its broad applicability and efficiency, and ImageBind achieving SOTA performance in zero-shot recognition across various modalities. To evaluate the effectiveness of BeFA, we applied it to several representative multimodal recommendation models, including LATTICE \cite{lattice}, BM3 \cite{bm3}, FREEDOM \cite{FREEDOM}, and MGCN \cite{mgcn}. Additionally, we compared BeFA with existing efficient parameter tuning methods, including LoRA \cite{lora} and Soft-Prompt Tuning \cite{softprompt}.  
\subsubsection{Evaluation Protocols \& Implementation}
Due to the length of the article, the details are shown in the Appendix.

\subsection{Overall performance}
\begin{figure}[t]
   \centering
   \includegraphics[width=1.0\linewidth]{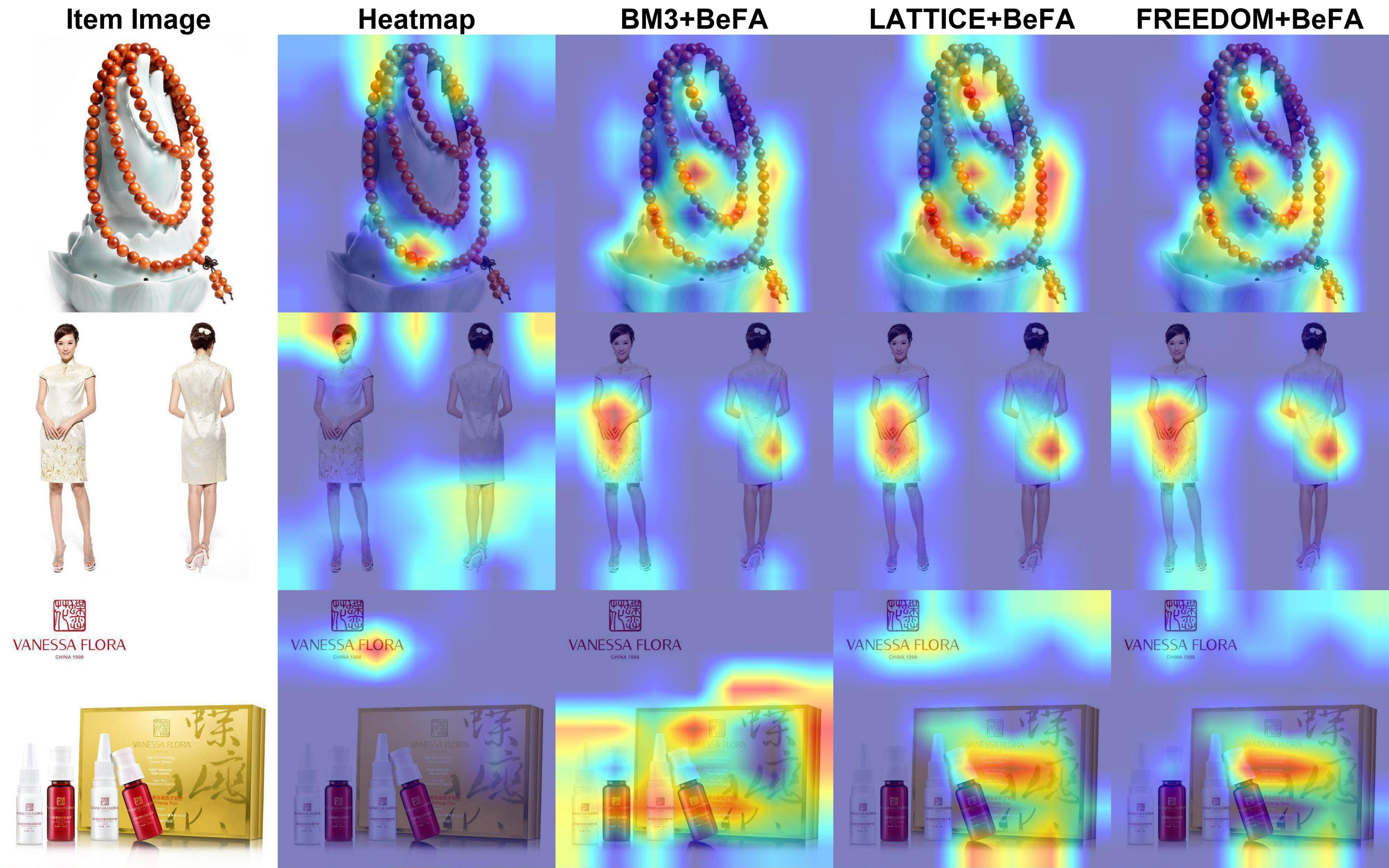} 
    \caption{Visualization analysis of the effect of adapter on feature purification. The shade of the colour represents the amount of attention weight.}
    \label{heatmap}
\end{figure}
\begin{table*}[t]
    \centering
    \small
    \setlength{\tabcolsep}{1.45mm} 
    \setlength{\abovecaptionskip}{0.05cm}

    \begin{tabular*}{\linewidth}{l|llll|llll|llll}
    \toprule
   \multicolumn{1}{c|}{}                            & \multicolumn{4}{c|}{TMALL}                                     & \multicolumn{4}{c|}{Microlens}                                                          & \multicolumn{4}{c}{H\&M}                                                                                              \\
  \multicolumn{1}{c|}{\multirow{-2}{*}{Dadasets}}  & {\color[HTML]{467886} R@10} & {\color[HTML]{467886} R@20} & {\color[HTML]{467886} N@10} & {\color[HTML]{467886} N@20} & {\color[HTML]{467886} R@10} & {\color[HTML]{467886} R@20} & {\color[HTML]{467886} N@10} & {\color[HTML]{467886} N@20} & {\color[HTML]{467886} R@10} & {\color[HTML]{467886} R@20} & {\color[HTML]{467886} N@10} & {\color[HTML]{467886} N@20} \\ \hline
  BM3                                              & 0.0189                      & 0.0298                      & 0.0102                      & 0.0132                      & 0.0510                      & 0.0851                      & 0.0278                      & 0.0375                      & 0.0204                & 0.0320                & 0.0114                      & 0.0144                \\
  BM3+LoRA                                          & \textbf{0.0215}             & \textbf{0.0333}             & \textbf{0.0117}             & \textbf{0.0150}            & 0.0510                      & 0.0850                      & 0.0280                      & 0.0376                     & 0.0191                      & 0.0290                      & 0.0110                      & 0.0135                      \\
  BM3+SoftPrompt                                   & 0.0193                      & 0.0298                      & 0.0102                      & 0.0131                      & 0.0517                & 0.0861                & 0.0285                & 0.0382                & 0.0202                      & 0.0311                      & 0.0114                & 0.0142                      \\
  BM3+BeFA                                         & 0.0212                      & 0.0319                     &  0.0115                    & 0.0144              & \textbf{0.0566}             & \textbf{0.0911}             & \textbf{0.0314}             & \textbf{0.0412}             & \textbf{0.0266}             & \textbf{0.0391}             & \textbf{0.0157}             & \textbf{0.0190}             \\ \hline
  LATTICE                                          & 0.0238                      & 0.0356                      & 0.0134                      & 0.0167                      & 0.0553                & 0.0886                      &  0.0308                & 0.0402                & 0.0289                      & 0.0427                      & 0.0161                      & 0.0197                      \\
  LATTICE+LoRA                                     & 0.0254                      & 0.0384                      & 0.0147                      & 0.0183                & 0.0550                      & 0.0884                      & 0.0304                      & 0.0399                      & 0.0289                      & 0.0434                      & 0.0163                      & 0.0201                      \\
  LATTICE+SoftPrompt                               & \textbf{0.0266}             & 0.0389                & 0.0148                & 0.0182                      & 0.0539                      & 0.0889                & 0.0294                      & 0.0393                      & 0.0301                & 0.0459                & 0.0166                & 0.0207                \\
  LATTICE+BeFA                                     & 0.0260                & \textbf{0.0403}             & \textbf{0.0183}             & \textbf{0.0183}             & \textbf{0.0593}             & \textbf{0.0943}             & \textbf{0.0328}             & \textbf{0.0427}             & \textbf{0.0317}             & \textbf{0.0498}             & \textbf{0.0171}             & \textbf{0.0217}             \\ \hline
  FREEDOM                                          & 0.0212                      & 0.0340                      & 0.0113                      & 0.0148                      & 0.0474                      & 0.0774                      & 0.0262                      & 0.0348                      & 0.0348                      & 0.0526                      & 0.0188                      & 0.0234                      \\
  FREEDOM+LoRA                                     & 0.0197                      & 0.0323                      & 0.0106                      & 0.0141                      & 0.0476                & 0.0774                & 0.0264                & 0.0349                & 0.0352                      & 0.0533                      & 0.0190                      & 0.0237                      \\
  FREEDOM+SoftPrompt                               & 0.0215                &  0.0342                & 0.0118                & 0.0154                & 0.0465                      & 0.0769                      & 0.0258                      & 0.0345                      & 0.0406                & 0.0577                & 0.0224                & 0.0268                \\
  FREEDOM+BeFA                                     & \textbf{0.0243}             & \textbf{0.0364}             & \textbf{0.0131}             & \textbf{0.0164}             & \textbf{0.0503}             & \textbf{0.0814}             & \textbf{0.0279}             & \textbf{0.0368}             & \textbf{0.0409}             & \textbf{0.0583}             & \textbf{0.0226}             & \textbf{0.0271}             \\ \hline
  MGCN                                             & 0.0249                      & 0.0380                      & 0.0135                      & 0.0171                      & 0.0618                & 0.0972                &  0.0342                & 0.0442                & 0.0367                      & 0.0549                      & 0.0204                      & 0.0251                      \\
  MGCN+LoRA                                        & 0.0260                      & 0.0391                      & 0.0141                      & 0.0179                      & 0.0598                      & 0.0963                      & 0.0335                      & 0.0438                      & 0.0367                      & 0.0554                      & 0.0203                      & 0.0252                      \\
  MGCN+SoftPrompt                                  & 0.0260                & 0.0391                & \textbf{0.0144}             & \textbf{0.0180}             & 0.0597                      & 0.0955                      & 0.0334                      & 0.0435                      & 0.0375                & 0.0557                & 0.0207                & 0.0254                \\
  MGCN+BeFA                                        & \textbf{0.0261}             & \textbf{0.0395}             & 0.0142                & 0.0179                & \textbf{0.0630}             & \textbf{0.1000}             & \textbf{0.0351}             & \textbf{0.0456}             & \textbf{0.0405}             & \textbf{0.0594}             & \textbf{0.0225}             & \textbf{0.0274}             \\ \hline
  \rowcolor{gray!40} \textit{\textbf{Avg Improve}}                & \textbf{2.44\%}                      & \textbf{1.71\%}                      & \textbf{7.89\%}                      & \textbf{0.48\%}        & \textbf{6.08\%}                      & \textbf{4.98\%}                      & \textbf{6.25\%}                      & \textbf{5.67\%}                                   & \textbf{11.11\%}                     & \textbf{9.59\%}                      & \textbf{12.58\%}                     & \textbf{11.44\%}                     \\ 
   \bottomrule
    \end{tabular*}
    \caption{Performance Comparison  with other Efficient Parameter Adaptation methods. The best result is in boldface and the second best is underlined. The $t$-tests validate the significance of performance improvements with $p$-value $\leq$ 0.05.}
    \label{comparison}
\end{table*}
\noindent \textbf{The effectiveness of BeFA.}
Table \ref{performance} demonstrates that our adapter enhances recommendation performance across all three datasets. Specifically, our adapter has achieved average improvements of 9.07\% and 11.02\% over the baseline in terms of Recall@20 and NDCG@20. This suggests that the pre-extracted content features do have deficiencies that negatively impact recommendation performance. Our processing of content features effectively reduces modality noise. This helps the multimodal recommendation models to better utilize the modality information and thus improve the recommendation performance. Furthermore, Table \ref{comparison} shows the performance of our adapter is better than existing efficient parameter tuning methods, highlighting its superior suitability for recommendation tasks. This suggests that our adapter uses behavioral information to better capture user-preferred content features, thus improving the performance.

\noindent \textbf{The generalization of BeFA.}
From Table \ref{performance}, we observe that BeFA improves performance across various multimodal recommendation models, highlighting its effectiveness as a generic plugin for all multimodal recommendations. Existing multimedia recommendation methods ignore this problem, thus leading to suboptimal results. Enhancing content features by BeFA, can make the content features better reflect the users' preference information. Thus make the existing methods more accurately model the users' preference. Meanwhile,  BeFA enhances performance for all types of encoders, indicating thar pre-trained encoders are generally defective. It can also be found that although more advanced feature encoder (ImageBind) can be used to extract information more comprehensively. The extracted features still contain a large amount of irrelevant information, resulting in generally do not accurately reflect the user's preference.
\begin{table}[t]
    \setlength{\abovecaptionskip}{0.05cm}

\begin{tabular}{cl|ll}
\toprule
\multicolumn{1}{l}{Dataset}                    & Method         & \#Param. & Time/E \\ \hline
\multirow{4}{*}{TMALL}                         & BM3            & 9.45M  & 0.38s  \\
                                               & BM3+LoRA       & +4.10K  & +0.04s  \\
                                               & BM3+SoftPrompt & +0.13K  & +0.02s  \\
                                               & BM3+BeFA          & +0.20M  & +0.08s  \\ \hline
\multicolumn{1}{r}{\multirow{4}{*}{Microlens}} & BM3            & 18.36M & 0.98s  \\
\multicolumn{1}{r}{}                           & BM3+LoRA       & +4.10K & +0.07s  \\
\multicolumn{1}{r}{}                           & BM3+SoftPrompt & +0.13K & +0.04s  \\
\multicolumn{1}{r}{}                           & BM3+BeFA          & +0.20M & +0.15s  \\ \hline

\multirow{4}{*}{H\&M}                          & BM3            & 21.26M & 1.20s  \\
                                               & BM3+LoRA       & +4.10K & +0.04s  \\
                                               & BM3+SoftPrompt & +0.13K & +0.01s  \\
                                               & BM3+BeFA          & +0.20M & +0.12s  \\ 
\bottomrule
\end{tabular}
\caption{The training cost. \#Param: number of tunable parameters, Time/E: averaged training time for one epoch.}
\label{paramters}
\end{table}

\noindent \textbf{The efficiency of BeFA.}
From Table \ref{paramters}, we observe that our adapter is more complex compared to LoRA and Soft-Prompt Tuning. However, the increase in the number of parameters after using our adapter remains small, ranging from 0.93\% to 2.09\% of the overall recommender system's parameters. The increase in overall training time is only about 15\%, which is comparable with other existing methods. This suggests that although our adapter is more complex and introduces additional parameters, it does not impose a burden on the training process.The substantial improvement in performance of our adapter highlights its efficiency. It trades a small increase in parameters and training time for a improvement in recommendation performance.
\subsection{Visualization Analysis}
\begin{figure}[t]
  \centering
  \includegraphics[width=1.0\linewidth]{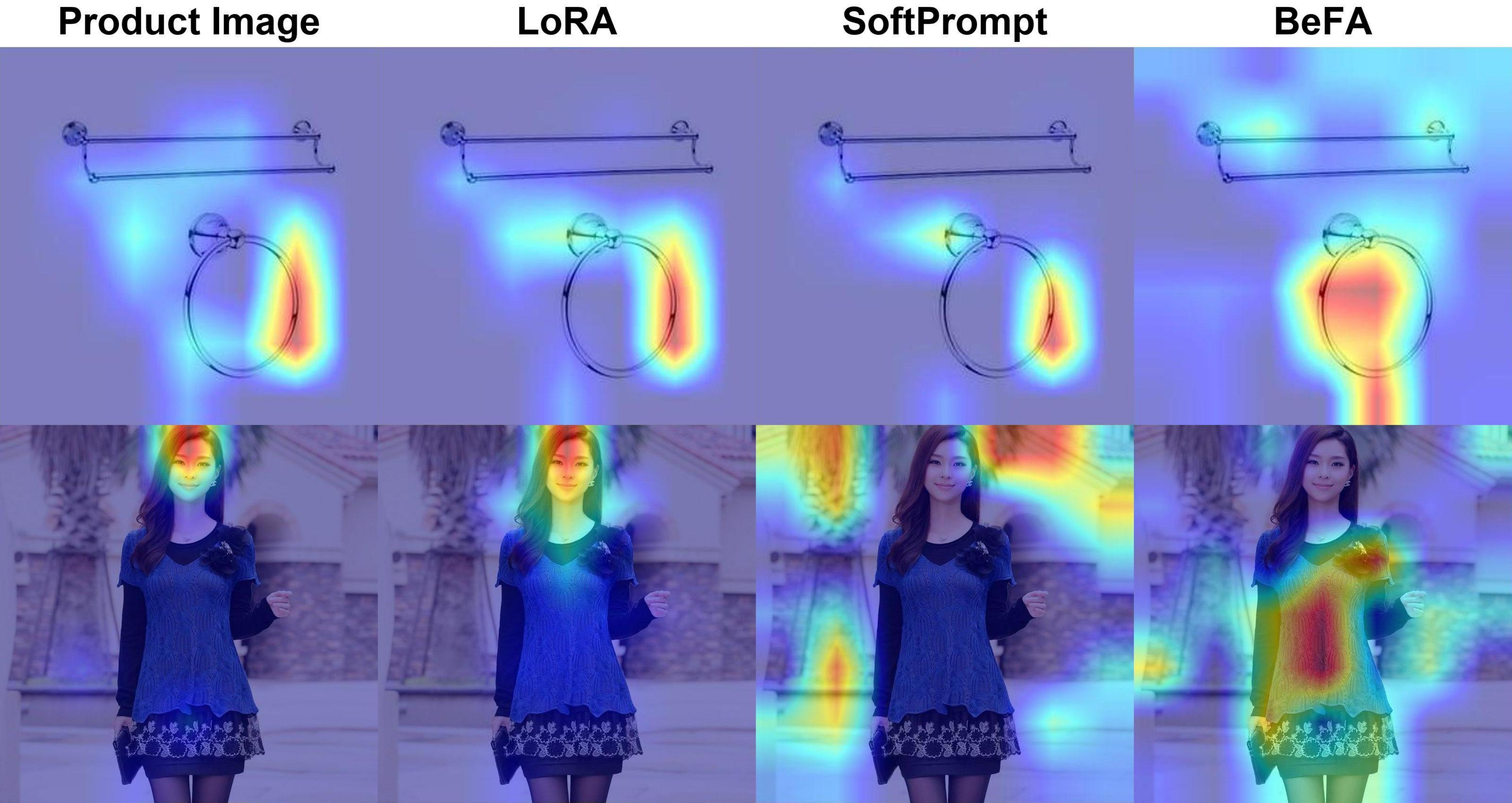}
  \caption{Comparison between different parameter-efficient adaptation methods.}
  \label{lora}
\end{figure}

\noindent \textbf{BeFA guides content features to more accurately reflect user preferences.} To visually demonstrate the  effect of our adapter on content features, we applied the purified features to the proposed visualization attribution analysis. As shown in Figure \ref{heatmap}, our adapter addresses issues of information drift and information omission. The adapted features focus more on the recommended items, with a marked decrease in attention to background elements unrelated to the recommendation.  Additionally, the adapted features more comprehensively capture item details. This comprehensive extraction of item details improves feature recognition, which enhances the recommender system's ability to model item features. Thus provides more relevant and accurate results when making recommendations. We also visualized the effect of BeFA on the distribution of visual representations. We find that the adapted features more accurately reflect the detailed information of the items, and the discriminability of the features is clearly improved. This is reflected in the distribution as a noticeable improvement in homogeneity \cite{pre}.

\noindent \textbf{BeFA is generalized with different encoders.} We utilize the content features extracted by two representative multimodal feature encoders, the ViT-32/B version of CLIP and ImageBind, and analyze the heatmaps of the content features within different downstream recommendation models after applying BeFA. Our analysis reveals that the content features extracted by both encoders exhibit issues of information drift and information omission. After applying BeFA, these content features are better reflect the recommended items themselves. This demonstrates  that the deficiencies of pre-trained feature encoders are common in multimodal recommendation tasks, while our adapter effectively adapts to the content features extracted by different feature encoders, highlighting its general applicability.
\begin{figure}[t]
  \centering
  \includegraphics[width=1.0\linewidth]{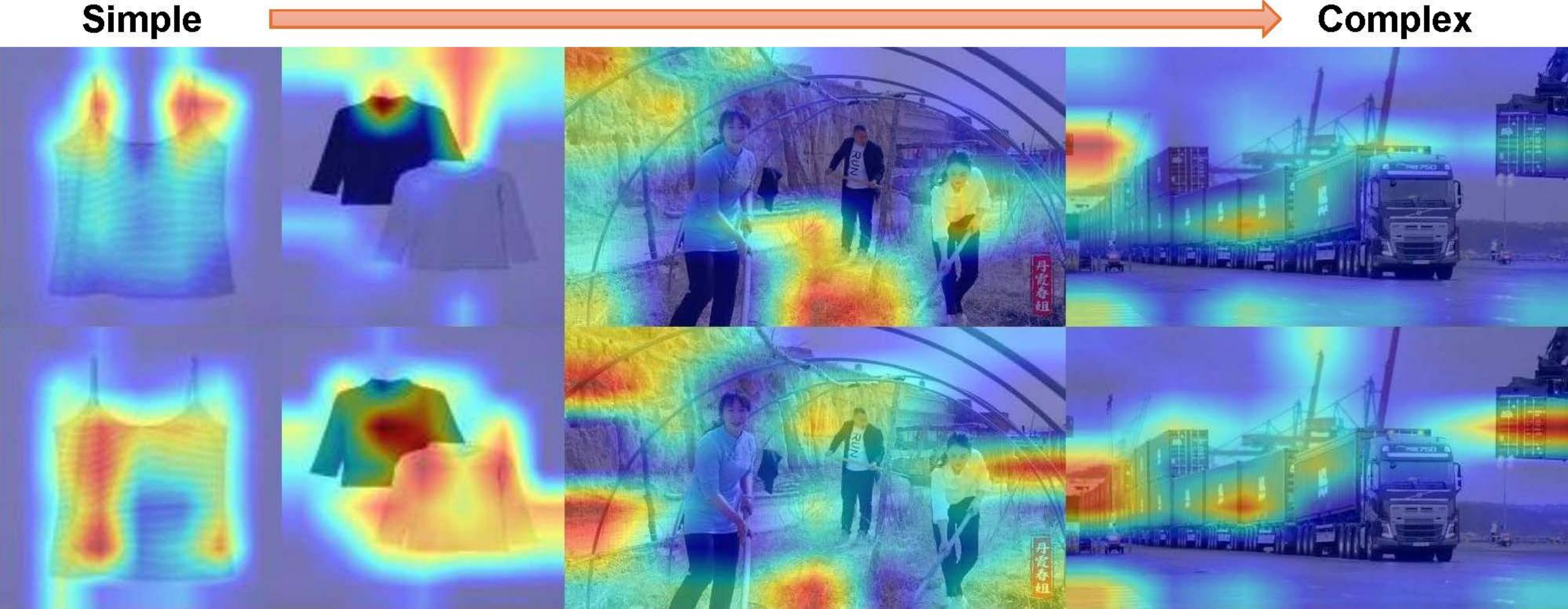}
  \caption{The effectiveness of BeFA on different scenarios. The top image is the heatmap and the bottom one is the heatmap after applying BeFA.}
  \label{HM_Microlens}
\end{figure}

\noindent \textbf{BeFA outperformance than other parameter tuning methods.} We visualize and analyze the content features adapted by LoRA and Soft-Prompt Tuning, as shown in Figure \ref{lora}. The results indicate that although LoRA and Soft-Prompt Tuning offer some optimization, the degree of adaptation is limited. These methods do not sufficiently address issues of information drift and omission, failing to focus accurately on the content of the item itself. This explains why existing adaptation methods perform poorly in recommendation tasks. In contrast, BeFA has a greater impact on content characterisation than existing methods. The quality of its adjusted content features is obviously higher. It can better model the items and thus improve the performance.

\noindent \textbf{BeFA performs well across various scenarios.} We employed three datasets ranging from simple to complex scenarios. The H\&M dataset comprises only item images with minimal interference, making it relatively clean. In contrast, the Microlens dataset comprises more complex images with more background interference. Our adapter demonstrated improvements across both datasets. Specifically, on the H\&M dataset, it increases Recall@20 and NDCG@20 by 14.17\% and 17.34\%. Meanwhile, on the Microlens dataset, our adapter increases Recall@20 and NDCG@20 by 4.58\% and 5.19\%. BeFA reduces the interference between content features and filters and retains the most relevant features for the recommendation task, as illustrated in Figure \ref{HM_Microlens}. As a result, it achieves better recommendation results even in various complex situations.
\subsection{Hyperparameter Analysis}
Due to the length of the article, specific analyses are provided in the Appendix.
\section{Conclusion}
In this paper, we introduce an attribution analysis for visualization analyzing deficiencies of the content features. We found that the content features suffer from information drift and information omission, which lead to a decrease in the performance of the recommender systems. To address these issues, we propose Behavior-driven Feature Adapter (BeFA) which refine content features through the guidance of behavioral information.
In our future work, we consider further designing more effective parameter-efficient adaptation methods.
\begin{appendix}
\section{Appendix}

\subsection{Theoretical Analysis}

In this section, we reveal the deficiencies of content features extracted by pre-trained feature encoders within the embedding space. Furthermore, we demonstrate that optimizing the consistency between the extracted features and the ideal features necessary for accurate recommendation predictions can significantly enhance recommendation performance. This analysis provides theoretical support for the effectiveness of our proposed adapter.

Analyzing features in the embedding space enables a better understanding and interpretation of their relationships. Formally, considering an item ${i}$ and its representation $\boldsymbol{e_{i,m}}$ in the embedding space, which represents the ideal content feature required by the recommendation model. On the basis of the item feature representation $\boldsymbol{e_{i,m}}$ in the embedding space, the recommender system calculates the probability distribution of the rating or clicking behavior of the user $u$ on the item $i$, which can be viewed as the prediction $P(r_{ui}|\boldsymbol{e_{i,m}},\boldsymbol{h_i},\boldsymbol{h_u})$.  Where $r_{ui}$ stands for the rating or click behavior of the user $u$ on item $i$. Our objective is to find the optimal content feature $\boldsymbol{e_{i,m}}$ to maximize the posterior distribution. This can be formally described by the  following objective function:
\begin{equation}
  L(\boldsymbol{e_{i,m}}) = \arg\max_{\boldsymbol{e_{i,m}}} P(r_{ui} | \boldsymbol{e_{i,m}},\boldsymbol{h_i},\boldsymbol{h_u})
\end{equation}
The representation $\boldsymbol{e_{i,m}'}$ extracted by the feature encoder can be seen as the prior distribution. However, there are deviations between the pre-extracted features and the ideal features. Let us denote this deviation as $\theta$.
\begin{equation}
  \theta = \arccos\left(\frac{\boldsymbol{e_{i,m}} \cdot \boldsymbol{e_{i,m}'}}{\|\boldsymbol{e_{i,m}}\| \|\boldsymbol{e_{i,m}'}\|}\right)
\end{equation}
A smaller $\theta$ indicates that the two representations are very close in the embedding space. This implies a high consistency between the pre-extracted features and the ideal features, which benefits the recommendation system in accurately capturing user interests and item similarities. By reducing $\theta$, the features become closer to the ideal features, thereby enhancing the performance of the recommendation system.

The part $p$ of the representation $\boldsymbol{e_{i,m}'}$ extracted by the encoder that is truly effective for the recommendation task can be represented as follows:
\begin{equation}
  p = \boldsymbol{e_{i,m}'} \cdot \left(\frac{\boldsymbol{e_{i,m}} \cdot \boldsymbol{e_{i,m}'}}{|\boldsymbol{e_{i,m}}| |\boldsymbol{e_{i,m}'}|} \right)=\boldsymbol{e_{i,m}'} \cdot \cos(\theta) 
\end{equation}
When the deviation $\theta$ is large, as $\boldsymbol{e_{i,m}'}$ in Figure \ref{angle}, content features exhibit the issues of information omission, causing the effective length of $p$ deviates significantly from $\boldsymbol{e_{i,m}}$. This results in a lack of crucial information in the extracted features, reflected as insufficient region of interest in the heatmap. Conversely, when $\theta$ is too large,as $\boldsymbol{e_{j,m}}$ in Figure \ref{angle}, content features exhibit the issues of information drift, with $p$ being located in the wrong quadrant and its effective direction opposing $\boldsymbol{e_{i,m}}$. This introduces a large amount of incorrect information, reflected as incorrect region of interest in the heatmap. In such circumstances, the recommendation system is easily to make incorrect recommendations, because the feature representation contradicts the users' preferences, failing to appropriately reflect the characteristics of items or users. In the recommendation task, we aim for a smaller deviation between $\boldsymbol{e_{i,m}'}$ and $\boldsymbol{e_{i,m}}$. To formalize this objective, we define an error function $D(\boldsymbol{e_{i,m}'}, \boldsymbol{e_{i,m}})$ to measure the deviation between the representations:
\begin{equation}
  D(\boldsymbol{e_{i,m}'},\boldsymbol{ e_{i,m}})= \left(1 - \frac{{\boldsymbol{e_{i,m}'} \cdot \boldsymbol{e_{i,m}}}}{{\|\boldsymbol{e_{i,m}'}\| \|\boldsymbol{e_{i,m}}\|}}\right)
\end{equation}
We seek to minimize the expected deviation measure $\Delta$. We can define $\Delta$ as:
\begin{equation}
  \Delta = \mathbb{E}_{P(\boldsymbol{e_{i,m}'})}[D(\boldsymbol{e_{i,m}'}, \boldsymbol{e_{i,m}})]
\end{equation}
A decline in the quality of $\boldsymbol{e_{i,m}'}$, indicated by an increase in the value of the error function $D(\boldsymbol{e_{i,m}'}, \boldsymbol{e_{i,m}})$, donates an augmented diversity between the representation extracted by the pre-trained encoder $\boldsymbol{e_{i,m}'}$ and the ideal representation $\boldsymbol{e_{i,m}}$. Consequently, this leads to an increase in the expected deviation $\Delta$, subsequently influencing the posterior distribution $P(r_{ui}|\boldsymbol{e_{i,m}},\boldsymbol{h_i},\boldsymbol{h_u})$. Such circumstances may impair the recommendation system's ability to accurately capturing the users' preferences, thereby reducing the accuracy of recommendations. By minimizing $\Delta$, we can ensure that the representation $\boldsymbol{e_{i,m}'}$ are closer to the ideal representation $\boldsymbol{e_{i,m}}$ in expectation. This adaptation of content features enhances the consistency with the ideal representation and thus improves the performance of the recommender system.
\begin{figure}[t]
  \appendix
 \setcounter{figure}{0}
 \setcounter{table}{0}
 \renewcommand{\thetable}{A\arabic{table}}
 \renewcommand{\thefigure}{A\arabic{figure}}
  \centering
  \includegraphics[width=\linewidth]{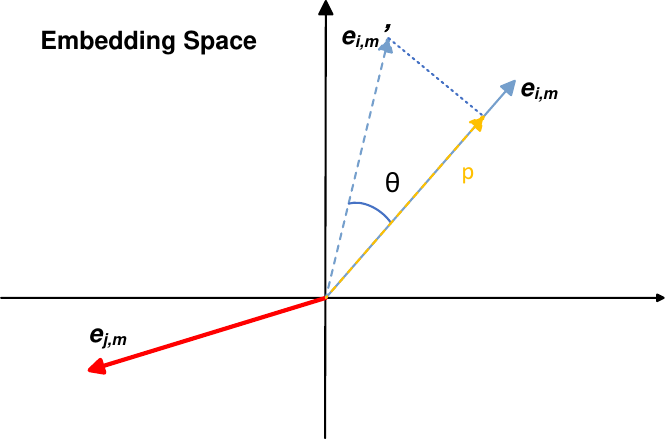}
  \caption{An illustration of a theoretical analysis of deficiency analysis}
  \label{angle}
\end{figure}
\subsection{Experiments}
\subsubsection{Datasets}
\begin{table}[h]
  \appendix
  \renewcommand{\thetable}{A\arabic{table}}

  \begin{tabular}{ccccl}
    \toprule
    Dataset & \#User & \#Item & \#Behavior  & Density \\
    \midrule
    TMALL & 13,104 & 7,848 & 151,928 & 0.148\% \\
    Microlens & 46,420 & 14,079 & 332,730 & 0.051\% \\
    H\&M & 43,543 & 16,915 & 369,945 & 0.050\% \\
  \bottomrule
\end{tabular}
  \caption{Statistics of the experimental datasets}
    \label{datasets}
\end{table}

We conducted experiments on three publicly available datasets: (a) {TMALL}\footnote{\url{https://tianchi.aliyun.com/dataset/140281}}; (b) {Microlens}\footnote{\url{https://recsys.westlake.edu.cn/MicroLens-50k-Dataset/}};  and (c) {H\&M}\footnote{\url{https://www.kaggle.com/datasets/odins0n/handm-dataset-128x128}}. All these three datasets are real-world datasets. We performed 10-core filtering on the raw data. The detailed information of the filtered data is presented in Table \ref{datasets}.
\subsubsection{Evaluation Protocols} 
For a fair comparison, we follow the evaluation settings in \cite{lattice,bm3} with the same 8:1:1 data splitting strategy  for training, validation and testing. Besides, we follow the all-ranking protocol to evaluate the top-K recommendation performance and report the average metrics for all users in the test set: $\mathrm{R@K}$ and $\mathrm{N@K}$, which are abbreviations for $\mathrm{Recall@K}$ \cite{recall} and $\mathrm{NDCG@k}$ \cite{NDCG}, respectively.
\subsubsection{Implementation}
We implement MMRec\footnote{\url{https://github.com/enoche/MMRec}} \cite{mmrec} based on PyTorch, which is a unified public repository designed for multimodal recommendation methods. To ensure fair comparison, we employed the Adam optimizer to optimize all methods and referred to the best hyperparameter settings reported in the original baseline paper. For general settings, we initialized embeddings with Xavier initialization of dimension 64, set the regularization coefficient to $\lambda_E = 10^{-4}$, and batch size to $B = 2048$. Early stopping and total epochs are fixed at 10 and 1000, respectively. We selecte the best model with the highest $\mathrm{Recall@20}$ metric on the validation set and reported metrics on the test set accordingly. Our experiments are done using RTX 4090 on windows system.
\subsubsection{Hyperparameter Analysis}
\begin{figure}[t]
\appendix
 \renewcommand{\thefigure}{A\arabic{figure}}
  \centering
  \includegraphics[width=\linewidth]{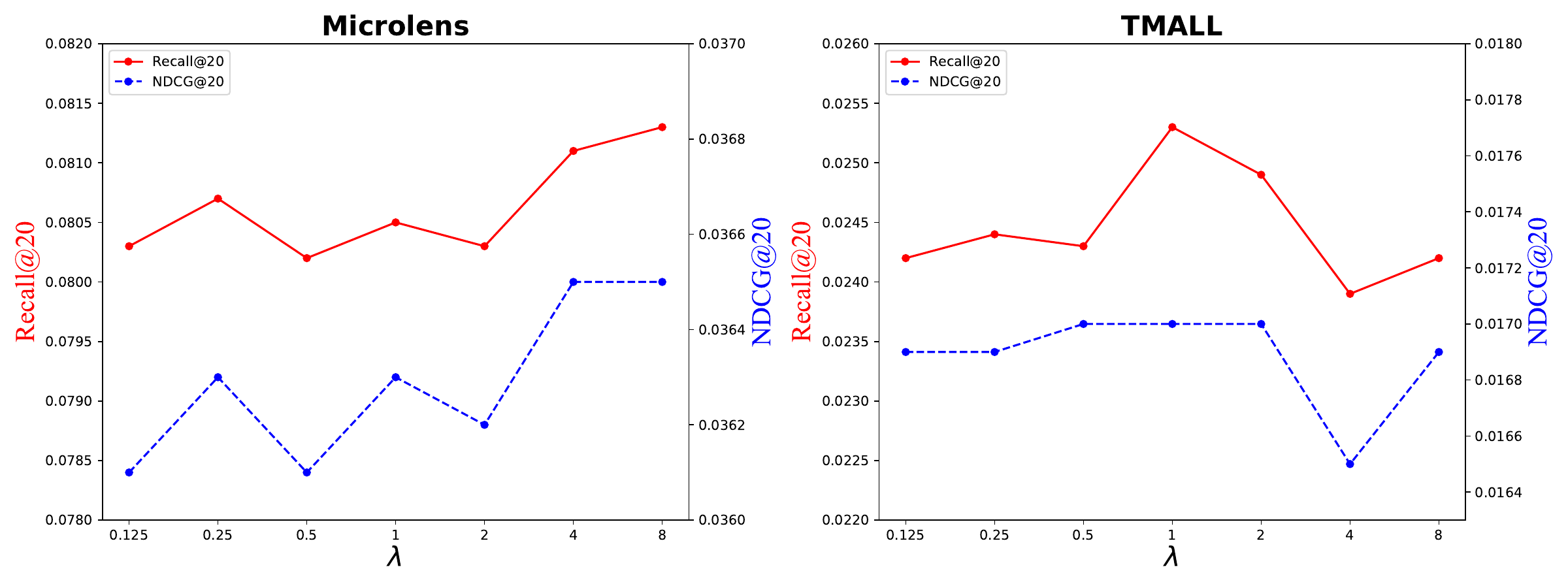}
  \caption{Performance comparison w.r.t. different size of the decoupling space relative to the original embedding space $\lambda$.}
  \label{hyper_fig}
\end{figure}

We investigate the size of the decoupling space relative to the original embedding space, as shown in Fig.\ref{hyper_fig}. The results show that the optimal size of the decoupling space is about four times the size of the original embedding space on some datasets. When the decoupling space dimension is small (e.g., 0.125x and 0.25x), it may create an information bottleneck,  leading to poor performance across all datasets in both $\mathrm{Recall@20}$ and $\mathrm{NCGD@20}$. A smaller decoupling space cannot hold sufficient information, making it difficult for the model to accurately capture and differentiate features, thereby negatively impacting recommendation performance.  Conversely, an excessively large decoupling space dimension may lead to dimensionality catastrophe, introducing noise and redundant features which complicates the model's ability to effectively distinguish and learn features in a high-dimensional space. Additionally, its number of parameters and the computational cost will increase dramatically, bringing additional burden to the training process, which is impractical for practical application scenarios.

Furthermore, we find that there is a significant difference in the performance of Microlens, TMALL and H\&M datasets under varying decoupling space sizes. For the Microlens dataset,the optimal performance is achieved when the decoupling space size is four times of the original embedding space. In contrast, in the TMALL dataset, although the overall trend is similar, the optimal point fluctuates slightly, suggesting that the size of the decoupling space may need to be adjusted on different datasets to obtain optimal results. The overall adapter performance fluctuates greatly with the change in the decoupling space size, highlighting the challenge of finding the optimal size.  This is one of the limitations of our work, which could be addressed by considering the decoupling space size as a learnable parameter in future research. This approach could adaptively adjust the decoupling space size for different application scenarios to achieve the best performance.

\end{appendix}
\section{Acknowledgment}
This work was supported the National Nature Science Foundation of China under Grants (No.62325206, 72074038), the Key Research and Development Program of Jiangsu Province under Grant BE2023016-4, the Natural Science Foundation of Jiangsu Province (BK.20210595) and the Postgraduate Research \& Practice Innovation Program of Jiangsu Province under Grant (KYCX23\_1026).
\nocite{*}
\bibliography{reference}

\end{document}